# Maximum Bipartite Matching Size and Application to Cuckoo Hashing

Yossi Kanizo, David Hay, and Isaac Keslassy

*Abstract*—Cuckoo hashing with a stash is a robust multiple choice hashing scheme with high memory utilization that can be used in many network device applications. Unfortunately, for memory loads beyond $0.5$, little is known on its performance.

In this paper, we analyze its average performance over such loads. We tackle this problem by recasting the problem as an analysis of the expected maximum matching size of a given random bipartite graph. We provide exact results for any finite system, and also deduce asymptotic results as the memory size increases. We further consider other variants of this problem, and finally evaluate the performance of our models on Internet backbone traces. More generally, our results give a tight lower bound on the size of the stash needed for any multiple-choice hashing scheme.

## I. INTRODUCTION

### A. Background

Network devices increasingly rely on hash tables to efficiently implement their algorithms, in fields as diverse as load-balancing, peer-to-peer, state management, monitoring, caching, routing, filtering, and security [1]–[6].

Because of the stringent *memory size constraints* in network devices, recent research is increasingly dealing with improving the memory efficiency of hash tables. In particular, *cuckoo hashing* has recently drawn a lot of attention due to its efficient space utilization along with its constant query and deletion times, as well as its constant expected insertion time (e.g., [7]–[18] and references therein). In cuckoo hashing, we want to insert $n$ elements into $m$ unit-sized bins. Each of the $n$ elements typically uses $d = 2$ independent hash functions, each pointing to an arbitrary bin. When an element arrives, it is placed in one of these 2 bins. If both bins are full, it displaces another element, which is then moved to the bin corresponding to its other choice. This process continues until all elements are placed, unless it is stopped and then the element cannot be inserted.

Cuckoo hashing is especially interesting because of its high memory utilization. In fact, consider the bipartite graph formed by the $n$ elements on one side, the $m$ bins on the other, and 2 links leaving each element for 2 bins according to the hash values of the element. Then the number of elements that cuckoo hashing inserts successfully *is exactly the size of the maximum matching* [11], [19], i.e. it is extremely sufficient. Past papers have in fact shown that up to a load $n/m = 0.5$, all

Y. Kanizo is with the Dept. of Computer Science, Technion, Haifa, Israel. Email: ykanizo@cs.technion.ac.il.

D. Hay is with the School of Computer Science And Engineering, Hebrew Univ.,Jerusalem, Israel. Email: dhay@cs.huji.ac.il

I. Keslassy is with the Dept. of Electrical Engineering, Technion, Haifa, Israel. Email: isaac@ee.technion.ac.il.

elements could fit in the hash table with high probability [7], [9]–[11].

Unfortunately, even this high efficiency of cuckoo hashing is not sufficient for network devices, because designers typically consider that at load 0.5, *half the memory is lost*. This is why memory-efficient hashing schemes attempt to pack even more elements by introducing an additional memory, called *stash* or overflow list, that stores a small number of elements outside the main hash table [6], [14]–[17]. The stash can then be implemented in hardware using for instance CAMs (content-addressable memories), which rely on associative memory and consume significantly more power [6], [18]. It can also rely on another hash table, or simply correspond to dropped packets in a lossy hash table [20].

This memory-efficient architecture with a stash enables the load of the hash-table to increase beyond $0.5$. Unfortunately, when the load gets beyond $0.5$, sizing the stash is not fully understood [18, Open Question 5]. *This paper is about analyzing this challenging case where the load exceeds* $0.5$.

To further understand why cuckoo hashing has a high utilization, it is important to notice that *online* cuckoo hashing with $d = 2$, succeeds in inserting an element if and only if an augmenting path originating from the corresponding vertex exists [11]. This is because inserting an element into a cuckoo hash table is equivalent to finding an augmenting path in the corresponding graph (that is, a path that starts from the vertex corresponding to the considered element, and alternates between unmatched and matched edges until it ends at a right-side vertex whose all edges are unmatched). Notice that for any sub-path $(r_1, v, r_2)$, where $v$ is a left-side vertex and $r_1, r_2$ are right-side vertices, $(r_1, v)$ must be a matched edge and $(v, r_2)$ an unmatched edge. Intuitively, this corresponds to moving element $v$ from bin $r_1$ to bin $r_2$. Since maximum size matching can be computed by finding such augmenting paths, when considering each left-side vertex only once and in arbitrary order, we can immediately conclude that the number of elements that a cuckoo hashing inserts successfully *is exactly the size of the maximum matching*. For example, all $n$ elements can be inserted if and only if the corresponding graph has a *perfect matching* (namely, a maximum matching of size $n$; see [11] for more details).

### B. Contributions

In this paper, we attempt to model the behavior of cuckoo hashing with a stash as the load gets beyond $0.5$. To do so, we essentially transform the problem into a problem in graph theory, then provide a theoretical analysis of its performance, and later evaluate its real-life behavior by using Internet backbone traces.

First, we study the *average performance* of cuckoo hashing by analyzing the *expected maximum matching size* in the bipartite graph introduced above. We decompose each random bipartite graph into *connected components*, and then separately analyze each component and evaluate the size of its local maximum bipartite match. The size of the maximum bipartite matching is the sum of the sizes of all local matches. Then, we count the number of connected components in the graph and thus derive the size of the maximum matching in the entire graph. Surprisingly, we can obtain an *exact expression* of the average performance of cuckoo hashing with a stash in any finite system.

We further show that the actual maximum matching size is sharply concentrated around its expected value. Thus, the difference between $n$ and the expected maximum matching size provides *the required size of the stash*, which should store all elements with high probability. To do so, we use concentration results based on applying Azuma's inequality to a Doob martingale, which is defined over the maximum matching size when exposing vertices one at a time. In practice, the goal of this result is *to help designers size their CAM stashes* by providing guarantees on the number of elements that can be inserted in the hash table with these stashes.

We next provide an exact analysis when the average number of choices is less than 2 to minimize the number of memory accesses. We further obtain a lower bound on the required stash size when the number of hashes $d$ exceeds 2. Our results for $d > 2$ rely on Huisimi tree enumerations. They illustrate the tradeoff between an improved memory efficiency and the need for more memory accesses, i.e. *the tradeoff between memory size and bandwidth*.

Finally, we evaluate cuckoo hashing with a stash on real-life Internet packet traces from an OC192 backbone link, using a 64-bit mix hash function. We show that when $n = m$, we can insert an average of $83.81\%$ of the packets within the hash table, and put the remainder in the CAM stash. Likewise, when $n = 0.6m$, that is, $20\%$ more than the threshold for perfect matching, we can insert in average $\approx 0.9938n$ of the packets, that is, the need stash size is $\approx 0.0062n$. We further confirm our analytical models and show that our bounds for $d > 2$ are typically within $1\%$ of the exact value.

Incidentally, our paper can lead to two interesting contributions. First, the paper analysis also provides exact results for the stash sizes when the numbers of elements $n$ and buckets $m$ are finite. This non-asymptotic analysis is particularly needed when $n$ and $m$ are known to be small. For example, as suggested in [5], the cuckoo hashing scheme can be used to store fingerprints of elements and thus enable set membership queries. To be able to move a fingerprint from one bucket to another, the hash value can only depend on the current location and on the fingerprint itself. This can be modeled by cuckoo hashing with a small finite number of elements and buckets, implying again that an asymptotic analysis cannot be applied in this case.

In addition, we note that for other multiple-choice hashing schemes, our results provide *a lower bound on the size of the stash*. This is because the maximum matching size of the graph is always an upper bound on the number of elements that can be inserted into the hash table. Moreover, since finding the maximum matching in bipartite graphs is a fundamental problem with a wide range of applications in computer science, we believe that our results have also a theoretical significance and may be used in other contexts.

*Paper Organization:* We start by surveying the relevant literature in Section II. Then we introduce the preliminary definitions in Section III. Section IV provides the expected maximum matching size of random bipartite graphs with left-side vertex degree 2, where a variation of the problem in which each left-side vertex degree is at most 2 is considered in Section V. Next, in Section VI, we solve the more appealing problem in which the right-side vertices are partitioned into two subsets, and each left-side vertex has exactly one edge to each of these subsets. In Section VIII we verify and evaluate our results, including by real-life trace-based experiments. Last, Section VII provides an upper bound on the expected maximum matching size when the constant left-side vertex degree is at least three. For the sake of readability, most of the proofs are presented in Appendix A.

## II. Related Work

Multiple-choice hashing schemes were first considered in the seminal paper of Azar et al. [21]. It showed that placing each element in the least occupied bin among a constant number $d$ of random bins significantly improves the maximum bin load to $\frac{\log \log n}{\log d} + O(1)$ with high probability (compared to the case where $d = 1$, in which the maximum bin load is $\log n \left(1 + O(1)\right)$). This result initiated an extensive research with many variants of multiple-choice hashing schemes, which typically exhibited the so-called *power of two random choices* with $d = 2$ [22]. For brevity, we next survey only works that directly correspond to our paper.

First, we relate to works which considered the same model as in this paper (a random bipartite graph with constant left-side vertex degree). Motivated by achieving a performance guarantee for the cuckoo hashing scheme [23], the main effort has been to find a load threshold, such that for any load below the threshold a perfect matching exists with high probability. It is known that a cuckoo hashing scheme with $d = 2$ succeeds with high probability if the load is less than a load threshold of 0.5, but fails when the load is larger than 0.5 [7]. Recent works [9]–[11] have settled the problem of finding the corresponding thresholds for $d > 2$. Another recent work [16], shows that cuckoo hashing with a stash of size $s$, $d = 2$, and a load factor less than 0.5 fails with probability $O\left(n^{-s}\right)$. Our paper differs in that we also consider load values beyond 0.5 for $d = 2$. Moreover, while most of the works investigate only the asymptotic behavior, we also present in our paper analytical expressions for finite random graphs along with the asymptotic ones.

The problem of finding the expected maximum matching size is also investigated assuming other models of random graphs, mainly trees. In [24] (and references therein) the authors investigate the expected maximum matching size of an $(r,s)$-tree, finding that for almost all $(n,n)$-trees the percentage of dark vertices in a maximum matching is at least 72%. A more recent work [25] presents results related to the expected maximum matching size of the class of simply-generated trees. A model of a loop graph is considered by [26], showing a lower bound on the expected maximum matching size. While using the cavity method of statistical physics [27], the authors find analytically the value under consideration for the Erdös graph $G(n, c/(n-1))$, where $c < 2.7183$. Our paper differs in that it considers a different model of random bipartite graphs, where each vertex in $L$ chooses a constant number of vertices in $R$.

Additional related works deal with the probability of a perfect matching in other random graph models. For instance, in a random directed bipartite graph with $n$ left-side and $n$ right-side vertices, and an outward degree $d$ at each vertex, the probability that the random bipartite graph contains a perfect matching approaches 1 if $d > 1$, but approaches 0 otherwise [28]. Also, in a random bipartite graph with $n$ left-side vertices, $n$ right-side vertices, $cn$ edges picked uniformly at random, and a degree of at least 2, there is a perfect matching with high probability [29].

Finally, conjectures in [30], [31] consider the expected minimum matching weight given a full bipartite graph with random exponentially distributed edge weights. These conjectures are proved in [32], [33].

## III. DEFINITIONS AND PROBLEM STATEMENT

Given two disjoint sets of vertices $L$ and $R$ of size $n$ and $m$ respectively, we consider a random bipartite graph $G = \langle L + R, E \rangle$, where each vertex $v \in L$ has $d = 2$ outgoing edges whose destinations are chosen independently at random among all vertices in $R$. We allow both choices to be the same vertex, implying that $G$ might have parallel edges. For brevity, we sometimes say that $v \in L$ *chooses* a vertex $v' \in R$ if $(v, v')$ is in $E$. The *load* of $G$ is denoted by $\alpha = \frac{n}{m}$.

We further consider cases when the *averages number of choices* is less than 2.

*Definition 1:* Let $d_v$ be the number of choices of each vertex $v \in L$. The *average number of choices* $a$ is the average left-side vertex degree, i.e. $a = \frac{\mathbf{E}(\sum_{v \in L} d_v)}{n} = \frac{\sum_{v \in L} \mathbf{E}(d_v)}{n}$.

First, in the deterministic case, we find the expected maximum matching size of the graph $G_a = \langle L + R, E \rangle$, where each vertex $v \in L$ independently chooses a predetermined number $d_v \in \{1, 2\}$ of random vertices in $R$, such that $a = \frac{d_1 + 2 \cdot d_2}{n}$.

Second, in the random case, we analyze the slightly different case of a random bipartite graph $G_p = \langle L + R, E \rangle$ where each vertex chooses two vertices with probability $p$ and one vertex with probability $1 - p$. This implies that in $G_p$, the average number of choices $a = 1 + p$.

Finally, we also consider a *static partitioning of the choices*; the set $R$ is partitioned into two disjoint sets $R_u$ and $R_d$ of sizes $\beta \cdot m$ and $(1 - \beta) m$. In that case, we consider a random bipartite graph $G_\beta = \langle L + (R_u \cup R_d), E \rangle$, where each vertex $v \in L$ chooses exactly one vertex in $R_u$ and another vertex in $R_d$.

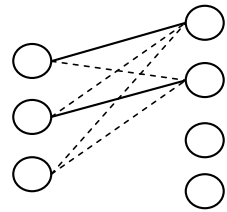

Fig. 1. An example bipartite graph with left-side vertex degree 2

This paper focuses on the expected size of the maximum size matching of $G$, which is captured by the following definition:

*Definition 2:* The operator $\mu(\cdot)$ extracts the *expected size of the maximum size matching*. It operates both on deterministic and random bipartite graphs. (Namely, for a deterministic graph $H$, $\mu(H)$ is simply the size of the maximum size matching of $H$.)

*Definition 3:* The *normalized limit expected maximum matching size* $\gamma = \lim_{n \to \infty} \frac{\mu(\cdot)}{n}$ is the limit percentage of the expected maximum matching size (out of the number of the vertices in $L$).

As shown in the literature, in real-world systems, practical hash functions usually work as if they were fully random [6]. Therefore, we model the hash functions in our theoretical analyses as such.

Our goal is to find both the *expected maximum matching size* as well as the *normalized limit expected maximum matching size* for the above-mentioned graph models.

## IV. EXPECTED CUCKOO PERFORMANCE: BIPARTITE GRAPHS WITH $d = 2$

We are now interested in evaluating the expected performance of cuckoo hashing with a stash. As explained in the Introduction, we approach the problem using a graph-theory perspective, since it is the same as evaluating the expected maximum matching size of the random bipartite graph $G$.

To do so, we consider the connected components of the random bipartite graph $G$. We start by stating some lemmas on these connected components, before establishing our main result on the expected matching size.

### A. Expected Maximum Matching Size

We now deal with a random graph, in which each left-side vertex chooses $d = 2$ right-side vertices (parallel edges are allowed). Note that further evaluation of the results reported here appears in Section VIII.

First, we quote a few useful lemmas (proved in [19]), before stating our result. As stated, the following lemmas are for a given bipartite graph $H = \langle L_H + R_H, E_H \rangle$, where each vertex in $L_H$ has degree 2 (parallel edges are allowed), with $|L_H| = s$ and $|R_H| = q$. An example bipartite graph $s = 3$, $q = 4$, and left-side vertex degree 2, appears in Figure 1. Dashed lines represent edges not in the maximum size matching, while solid lines represent edges in the maximum size matching.

*Lemma 1:* If $s \leq q - 2$, then $H$ is not connected.

*Lemma 2:* If $H$ is connected and $s \geq q$, then $\mu(H) = q$.

*Lemma 3:* If $H$ is connected and $s = q-1$ then $\mu(H) = s$.

*Lemma 4:* For any graph with $s = q-1$, $H$ is connected if and only if it is a tree.

*Lemma 5:* The number $T_s$ of connected bipartite graphs $H$ whose $|L_H| = s$ and $|R_H| = s+1$ is $T_s = (s+1)^{s-1} s!$.

We can now prove the next theorem on our bipartite graph $G$, which is the main result of this paper. We remind that this theorem states the expected number of elements $\mu(G)$ that can be inserted by our cuckoo hashing scheme with a stash. Therefore, $n - \mu(G)$ also gives us the *expected stash size*.

*Theorem 1:* Let $d = 2$ and $b = \min\{n, m-1\}$. The expected maximum matching size $\mu(G)$ is

$$\mu(G) = m - \sum_{s=0}^{b} \binom{n}{s}\binom{m}{s+1} \times \left(1 - \frac{s+1}{m}\right)^{2(n-s)} \left(\frac{s+1}{m}\right)^{2s} \frac{2^s s!}{(s+1)^{s+1}}.$$

*Proof:* Let $M$ be a maximum matching of $G$. Our proof is based on counting the expected number of vertices in $R$ that are not part of $M$, and on the decomposition of $G$ into its connected components.

Lemma 1 yields that any connected component of $G$ with $s$ left-side vertices has at most $s+1$ right-side vertices. We call a connected component with $s$ left-side vertices and $s+1$ right-side vertices *a deficit component of size $s$*. Lemma 3 implies that the maximum matching size of any such deficit component is $s$. Therefore, *exactly one* of its right-side vertices is not part of $M$. Notice that in all other connected components, where $q < s+1$, the maximum matching size of $G$ is exactly $q$ (Lemma 2), implying that all their right-side vertices are part of $M$.

Thus, in order to calculate the size of $M$, it suffices to count *the number of deficit components $x$*. The size of $M$ is $m - x$ because exactly $x$ right-side vertices do not participate in $M$, one for each deficit component.

Let $P_s = \frac{2^s T_s}{(s+1)^{2s}}$ be the probability that a bipartite graph $H = \langle L_H + R_H, E_H \rangle$ is connected, with degree 2 for all vertices in $L_H$, where $|L_H| = s$ and $|R_H| = s+1$.

The expected number of deficit components of size $s$ is $\binom{n}{s}\binom{m}{s+1} \cdot \left(1 - \frac{s+1}{m}\right)^{2(n-s)} \cdot \left(\frac{s+1}{m}\right)^{2s} \cdot P_s$. The above expression consists of the following factors (in order):

(i) choosing the $s$ vertices in $L$;
(ii) choosing the $s+1$ vertices in $R$;
(iii) the probability that all $s+1$ vertices in $R$ may be connected only to the chosen $s$ vertices in $L$;
(iv) the probability that all $s$ vertices in $L$ are only connected to the $s+1$ vertices in the right side; and,
(v) the probability that all chosen vertices are connected.

Finally, we calculate $x$ by summing over all possible values on $s$. As mentioned before, the expected size of $M$ is given by $m - x$. We get: $\mu(G) = m - \sum_{s=0}^{b} \binom{n}{s}\binom{m}{s+1} \cdot \left(1 - \frac{s+1}{m}\right)^{2(n-s)} \cdot \left(\frac{s+1}{m}\right)^{2s} \cdot P_s$, where $b = \min\{n, m-1\}$, $P_s = \frac{2^s T_s}{(s+1)^{2s}}$, and $T_s = (s+1)^{s-2} \cdot (s+1)!$, as found in Lemma 5. ∎

### B. Concentration Result

We next show that the size of the maximum matching is highly concentrated around its expectation $\mu(G)$. In other words, this means that our stash occupancy will be close to its average value, which can help us *size the stash* more accurately by providing performance guarantees on its performance.

In order to prove this result, we apply Azuma's inequality to a Doob martingale (more specifically, the martingale is a vertex exposure martingale of the left-side vertices).

Note that as long as all left-side vertices pick their edges independently, this concentration result holds regardless of the value of $d$, and more generally regardless of the specific distribution over which the hash functions are defined. Therefore, the concentration result applies also for the settings of Sections V – VII.

*Theorem 2:* Let $H$ be a specific instance of the random graph $G$, as defined in Section III. For any $\lambda > 0$, $\Pr(|\mu(H) - \mu(G)| > \lambda\sqrt{n}) < 2e^{-\lambda^2/2}$.

*Proof:* Our notations follow those of [34]. We first define an exposure martingale, which exposes one left-side vertex at a time, along with all its outgoing edges. This martingale is equivalent to a regular vertex exposure martingale, in which all right-side vertices are exposed first, and then left-side vertices are exposed one by one.

Specifically, let $G$ be the probability space of all two-choice bipartite graphs as defined in Section III and $f$ the size of the maximum size matching of a specific instance. Assume an arbitrary order of the left-side vertices $L = \{v_1, \ldots v_n\}$, and define $X_0, \ldots, X_n$ by $X_i(H) = E[f(G) \mid \forall x \leq i, \forall v_y \in R, (v_x, v_y) \in G \text{ iff } (v_x, v_y) \in H]$. Note that $X_0(H) = \mu(G)$ since no edges were exposed, while $X_n(H) = \mu(H)$ as all edges are exposed.

Clearly, $f$ satisfies the vertex Lipschitz condition since if two graphs $H$ and $H'$ differ at only one left-side vertex, $|f(H) - f(H')| \leq 1$ (either that vertex is in the maximum matching or not). Thus, since each left-side vertex makes independent choices, [34, Theorem 7.2.3] implies that the corresponding vertex exposure martingale satisfies $|X_{i+1} - X_i| \leq 1$. Hence, by applying Azuma's inequality, we immediately get the concentration result. ∎

Notice that if we are interested only in one-sided bounds, we can get a slightly tighter result: $\Pr(\mu(G) - \mu(H) > \lambda\sqrt{n}) < e^{-\lambda^2/2}$. This is exploited in the following corollary, which shows that to obtain a given overflow fraction, the needed stash size grows sub-linearly with $n$ beyond its average value.

*Corollary 3:* To achieve an overflow fraction of $\epsilon$ in cuckoo hashing with stash, when inserting $n$ elements to $m$ bins, a stash of size $n - \mu(G) + \sqrt{2n \cdot \ln(1/\epsilon)}$ suffices, where $\mu(G)$ is defined in Theorem 1.

*Proof:* If a stash of size $n - \mu(G) + \sqrt{2n \cdot \ln 1/\epsilon}$ is used, cuckoo hashing fails if and only if $n - \mu(H) > n - \mu(G) + \sqrt{2n \cdot \ln 1/\epsilon}$, or by rewriting it, $\mu(G) - \mu(H) > \sqrt{2n \cdot \ln 1/\epsilon}$. By substituting $\lambda = \sqrt{2 \cdot \ln 1/\epsilon}$ in the above one-sided bound, we get the claimed result. ∎

### C. Limit Normalized Expected Maximum Matching Size

We are now interested in the asymptotic expression where $n \to \infty$ with $\alpha = \frac{n}{m}$ constant. The following results show an interesting connection between the limit normalized expected maximum matching size and the Lambert-$W$ function, and even a connection between the perfect matching threshold and the radius of convergence of the Lambert-$W$ function [19], [35].

For further details on the Lambert-$W$ function, see also Appendix B.

*Theorem 4:* Let $d = 2$. The limit normalized expected maximum matching size $\gamma = \lim_{n \to \infty} \frac{\mu(G)}{n}$ is given by:

$$\gamma = \frac{1}{\alpha} + \frac{1}{2\alpha^2} \cdot W\left(-2\alpha \cdot e^{-2\alpha}\right) + \frac{1}{4\alpha^2} W^2 \left(-2\alpha \cdot e^{-2\alpha}\right), \quad (1)$$

where the Lambert-$W$ function is the inverse function of the function $\omega(x) = xe^x$.

*Proof:* We compute the limit of $\frac{\mu(G)}{n}$ as $n \to \infty$ such that $\alpha = \frac{n}{m}$:

$$\gamma = \lim_{n \to \infty} \frac{1}{n} \left( m - \sum_{s=0}^{b} \binom{n}{s}\binom{m}{s+1} \times \left(1 - \frac{s+1}{m}\right)^{2(n-s)} \left(\frac{s+1}{m}\right)^{2s} P_s \right)$$

We find through differentiation that $\left(1 - \frac{s+1}{m}\right)^{2(n-s)}$ is an increasing function with respect to $n$ (where $m = \frac{n}{\alpha}$). Moreover, the expansion of $\frac{1}{n} \cdot \binom{n}{s} \binom{m}{s+1} \cdot \left(\frac{s+1}{m}\right)^{2s}$ shows that it is also an increasing function. Therefore, their product is also increasing and, by the monotone convergence theorem [36], we get

$$\gamma = \frac{m}{n} - \sum_{s=0}^{b} \lim_{n \to \infty} \left( \frac{1}{n} \cdot \binom{n}{s}\binom{m}{s+1} \times \left(1 - \frac{s+1}{m}\right)^{2(n-s)} \cdot \left(\frac{s+1}{m}\right)^{2s} \cdot P_s \right)$$

By substituting the expression for $P_s$, and using the facts that $\binom{n}{s} = \frac{n^s}{s!} + O\left(n^{s-1}\right)$ and $\lim_{n \to \infty} (1 + a/n)^n = e^a$, we deduce:

$$\gamma = \frac{m}{n} - \frac{1}{n} \sum_{s=0}^{\infty} \frac{n^s}{s!} \frac{m^{s+1}}{(s+1)!} e^{-2\alpha(s+1)} \cdot \frac{(s+1)^{2s}}{m^{2s}} \cdot \frac{2^s (s+1)^{s-1} s!}{(s+1)^{2s}}$$

By substituting $m = \frac{n}{\alpha}$, and simplifying the above expression, we get:

$$\gamma = \frac{1}{\alpha} - \frac{1}{\alpha} \cdot \sum_{s=0}^{\infty} \alpha^s \cdot 2^s \cdot \frac{(s+1)^{s-1}}{(s+1)!} \cdot e^{-2\alpha(s+1)}$$

$$= \frac{1}{\alpha} - \frac{1}{2\alpha^2} \cdot \sum_{j=1}^{\infty} \left(-2\alpha \cdot e^{-2\alpha}\right)^j \cdot \frac{(-j)^{j-2}}{j!}$$

Let $T(x) = \sum_{j=1}^{\infty} \frac{(-j)^{j-2}}{j!} \cdot x^j$ be a formal power series, where by substituting $x = -2\alpha \cdot e^{-2\alpha}$ we get the above expression. By differentiating $T(x)$ and multiplying by $x$, we get:

$$x \cdot \frac{d}{dx} T(x) = -\sum_{j=1}^{\infty} \frac{(-j)^{j-1}}{j!} \cdot x^j = -W(x),$$

where the Lambert-$W$ function is the inverse function of the function $\omega(x) = xe^x$ [35], and the last equality follows from its known Taylor expansion that converges as long as $x$ is within the radius of convergence with $|x| \leq e^{-1}$ [35].

Given that $x \cdot \frac{d}{dx} T(x) = -W(x)$, we compute $T(x)$:

$$T(x) = \int \frac{1}{x} \cdot (-W(x)) \, dx = -W(x) - \frac{1}{2} W^2(x),$$

with convergence within $|x| \leq e^{-1}$.

Interestingly, the function $f(\alpha) = -2\alpha \cdot e^{-2\alpha}$ gets its minimum at $\alpha = 0.5$, where it precisely equals the radius of convergence $-e^{-1}$. Therefore, for all $\alpha$ we can substitute $x = -2\alpha \cdot e^{-2\alpha}$, since we are within the radius of convergence of $T(x)$, and we finally derive the result. ∎

We note that this particular asymptotic result can be also achieved by the theory of giant components in random graphs [34], [37]. However, this technique is not applicable for finite $n$ and $m$, and cannot be used to derive most of the other results in this paper. (A proof using this technique appears in the appendices).

The following corollary shows that for $\alpha = \frac{n}{m} \leq \frac{1}{2}$, the probability for a right-side vertex to be part of a maximum matching goes to 1. This corollary also follows from the previously known result that there is a perfect matching with high probability in cuckoo hash tables with load $\alpha \leq \frac{1}{2}$ [7].

*Corollary 5:* Let $d = 2$ and $\alpha = \frac{n}{m} \leq \frac{1}{2}$. Then the limit normalized expected maximum matching size is $\gamma = \lim_{n \to \infty} \frac{\mu(G)}{n} = 1$.

*Proof:* In case $\alpha \leq \frac{1}{2}$, $W\left(-2\alpha \cdot e^{-2\alpha}\right)$ equals $-2\alpha$, thus, $\gamma = \frac{1}{\alpha} + \frac{1}{2\alpha^2} \cdot (-2\alpha) + \frac{1}{4\alpha^2} (-2\alpha)^2 = 1$ ∎

## V. Cuckoo with Low Memory Bandwidth: Bipartite Graphs With $d_v \leq 2$

In this section we are interested in a *low-memory-bandwidth version* of the cuckoo hash algorithm. We now let each element choose either 1 or 2 bins instead of only 2 bins, to force them to access less bins and use less memory I/O bandwidth.

Formally, we relax the constraint that each vertex in $L$ chooses exactly 2 vertices in $R$, and let each left-side vertex choose either 1 or 2 right-side vertices. Since we can divide the set of vertices either deterministically or randomly, we will discuss the results in both cases. These results correspond for example to cases in which the average number of choices, as defined below, is important (e.g. [15]). See also [38] for a similar model.

Note that further evaluation of the results reported in this section can be found in Section VIII-B.

### A. Connected Components in Deterministic Graphs

As in Section IV-A, we now consider a deterministic bipartite graph $H = \langle L_H + R_H, E_H \rangle$, with $|L_H| = s$ and $|R_H| = q$. We assume that the degree of each vertex in $L_H$ is at most 2.

*Proposition 1:* Lemmas 1, 2, and 3 hold also when the degree of each vertex in $L_H$ is at most (but not necessarily) 2.

Note that the proofs remain almost identical to the original proofs, replacing a few equalities with the corresponding inequalities.

*Lemma 6:* Let $s+1 = q$. If $H$ is connected then the degree of each vertex in $L_H$ is 2.

### B. Expected Maximum Matching Size

*1) Predetermined Number of Choices:* In this section, we assume that each vertex $v \in L$ independently chooses $1 \leq d_v \leq 2$ random vertices in $R$, where $d_v$ is predetermined. The following result provides the expected maximum matching size in this case.

*Theorem 6:* Given a predetermined average number of choices $a$, let $d_1 = (2-a) \cdot n$ and $d_2 = n - d_1 = (a-1) \cdot n$ be the number of vertices in $L$ that choose one and two vertices in $R$, respectively. The expected maximum matching size $\mu(G_a)$ is given by:

$$\mu(G_a) = m - \sum_{s=0}^{b} \binom{d_2}{s} \binom{m}{s+1} \times \left(1 - \frac{s+1}{m}\right)^{2(d_2-s)+d_1} \left(\frac{s+1}{m}\right)^{2s} \frac{2^s \cdot s!}{(s+1)^{s+1}},$$

where $b = \min\{d_2, m-1\}$.

*2) Random Number of Choices:* In this section, we assume that each vertex $v \in L$ independently chooses $1 \leq d_v \leq 2$ random vertices in $R$, where for each $v \in L$, $d_v$ equals 2 with probability $p$, and it equals 1 with probability $1-p$. Based on Theorem 6, the following result reflects the expected maximum matching size in this case.

*Theorem 7:* The expected maximum matching size $\mu(G_p)$ is given by
$\mu(G_p) = \sum_{d_2=0}^{n} \binom{n}{d_2} \cdot p^{d_2} \cdot (1-p)^{n-d_2} \cdot \mu\left(G_{a=1+\frac{d_2}{n}}\right)$, where $\mu(G_a)$ is given by Theorem 6.

### C. Limit Normalized Expected Maximum Matching Size

*1) Predetermined Number of Choices:* We are also interested in the asymptotic expression, where $n \to \infty$, such that we fix both the load $\alpha = \frac{n}{m}$ and the average number of choices $a = \frac{d_1 + 2 \cdot d_2}{n}$ of the vertices. This is reflected in the following theorem.

*Theorem 8:* The limit normalized expected maximum matching size $\gamma_a = \lim_{n \to \infty} \frac{\mu(G_a)}{n}$ with average number of choices $a \in (1,2]$ is given by: $\gamma_a = \lim_{n \to \infty} \frac{\mu(G_a)}{n} = \frac{1}{\alpha} + \frac{W(-2\alpha(a-1) \cdot e^{-a\alpha})}{2\alpha^2 \cdot (a-1)} + \frac{W^2(-2\alpha(a-1) \cdot e^{-a\alpha})}{4\alpha^2 \cdot (a-1)}$. For $a = 1$, it is given by $\gamma_a = \lim_{n \to \infty} \frac{\mu(G_a)}{n} = \frac{1}{\alpha} - \frac{1}{\alpha} \cdot e^{-\alpha}$.

Interestingly, if even a small fraction of the elements do not have choice then the expected maximum matching size is not 1. This is reflected in the following corollary.

*Corollary 9 ((No) Perfect Matching):* If $1 \leq a < 2$ then $\gamma_a < 1$.

*2) Random Number of Choices:* We now study the case of the random bipartite graph $G_p = \langle L+R, E\rangle$, where each vertex chooses two vertices with probability $p$ (and one vertex with probability $1-p$). As we show in the next theorem, the asymptotic expression can be derived by $\gamma_a$.

*Theorem 10:* The limit expected maximum matching size $\gamma_p = \lim_{n \to \infty} \frac{\mu(G_p)}{n}$ where each vertex chooses two vertices with probability $p$ (and one vertex with probability $1-p$) is $\gamma_p = \gamma_{a=1+p}$.

## VI. SINGLE-PORTED CUCKOO: STATIC PARTITIONING OF THE CHOICES

We now consider a popular cuckoo-hashing implementation variant in which the bins are statically partitioned into two equal sets, and each element holds one hash function to each set. This variant is *easier to implement in hardware*, because it can be implemented using two simple *single-ported memories*, instead of a single dual-ported one.

Formally, we consider the random bipartite graph $G_\beta = \langle L + (R_u \cup R_d), E \rangle$, where $R$ is now partitioned into two disjoint subsets $R_u$ and $R_d$ with $|R_u| = \beta \cdot m$ and $|R_d| = (1-\beta)m$. Each vertex $v \in L$ independently chooses a single random vertex in $R_u$ and another single random vertex in $R_d$. This corresponds, for example, to a hashing scheme that selects non-overlapping sets of bins as images of its hash functions (e.g., as in multilevel hashing scheme [39] or d-left [40]).

Note that further evaluation of the results reported in this section can be found in Section VIII-C.

### A. Connected Components in Deterministic Graphs

The following lemma counts all the possible bipartite graphs $H_{ud}$ of the form $\langle L_H + (R_{H_u} \cup R_{H_d}), E_H \rangle$ with degree 2 for each vertex in $L_H$, where $|L_H| = s$, $|R_{H_u}| = i$ and $|R_{H_d}| = j$, such that each vertex $v \in L_H$ is connected using a single edge to some vertex in $R_{H_u}$ and another single edge to some vertex in $R_{H_d}$.

*Proposition 2:* Lemmas 1, 2, 3, and 4 hold for this case as well.

*Lemma 7:* Let $s = i + j - 1$. The number $T_{i,j}$ of connected bipartite graphs is $T_{ij} = i^{j-1} \cdot j^{i-1} \cdot s! = i^{j-1} \cdot j^{i-1} \cdot (i+j-1)!$

### B. Expected Maximum Matching Size

In the next theorem we find the expected maximum matching size with a static partition of the right-side vertices.

*Theorem 11:* Given the static partitioning of the bipartite graph $G_\beta$, the expected maximum matching size $\mu(G_\beta)$ is

$$\mu(G_\beta) = m - \sum_{s=0}^{n} \binom{n}{s} \sum_{i=b_1}^{b_2} \binom{\beta \cdot m}{i}\binom{(1-\beta)\cdot m}{s+1-i} \left(1 - \frac{i}{\beta \cdot m}\right)^{n-s} \times \left(1 - \frac{s+1-i}{(1-\beta)\cdot m}\right)^{n-s} \left(\frac{i}{\beta \cdot m}\right)^s \left(\frac{s+1-i}{(1-\beta)\cdot m}\right)^s \times P_{i,s+1-i},$$

where $b_1 = \max\{0, s+1-(1-\beta)\cdot m\}$, $b_2 = \min\{s+1, \beta \cdot m,\}$, $P_{ij} = \frac{T_{ij}}{(i \cdot j)^{i+j-1}}$, and $T_{ij} = i^{j-1} \cdot j^{i-1} \cdot (i+j-1)!$ (as given in Lemma 7).

*Proof:* Similarly to the proof of Theorem 1, our proof is based on counting the expected number of vertices in $L$ that are not in some specific maximum matching $M$ of $G_\beta$, based on the decomposition of $G$ into its connected components. As in the proof of Theorem 1, we consider the number of connected components with exactly $s$ vertices in $L$ and $q = s+1$ vertices in $R_u \cup R_d$, where we have to sum over all possible combinations $(i, s+1-i)$, where $i$ corresponds to the number of vertices taken from $R_u$ and $s+1-i$ corresponds to those taken from $R_d$.

Thus, the expected number of connected components in $G_\beta$ with $s$ vertices in $L$, $i$ vertices in $R_u$ and $s+1-i$ vertices in $R_d$ is given by:

$$\binom{n}{s} \cdot \binom{\beta \cdot m}{i}\binom{(1-\beta)\cdot m}{s+1-i} \cdot \left(1 - \frac{i}{\beta \cdot m}\right)^{n-s} \cdot$$
$$\left(1 - \frac{s+1-i}{(1-\beta)\cdot m}\right)^{n-s} \cdot \left(\frac{i}{\beta \cdot m}\right)^s \cdot \left(\frac{s+1-i}{(1-\beta)\cdot m}\right)^s \cdot P_{i,s+1-i},$$

The above expression consists of the following factors (in order):

*(i)* choosing the $s$ vertices in $L$;
*(ii)* choosing the $i$ vertices in $R_u$;
*(iii)* choosing the $s+1-i$ vertices in $R_d$;
*(iv)* the probability that all $i$ vertices in $R_u$ may be connected only to the chosen $s$ vertices in $L$;
*(v)* the probability that all $s+1-i$ vertices in $R_d$ may be connected only to the chosen $s$ vertices in $L$;
*(vi)* the probability that all $s$ vertices in $L$ are only connected to the $i$ vertices in $R_u$;
*(vii)* the probability that all $s$ vertices in $L$ are only connected to the $s+1-i$ vertices in $R_d$; and,
*(viii)* the probability that all chosen vertices are connected.

Finally, adding the expressions for all possible $s$'s and $i$'s and subtracting it from $m$ yields the claimed result. ∎

### C. Limit Normalized Expected Maximum Matching Size

As in the last sections, we are also interested in the asymptotic expression where $n \to \infty$ with both fixed $\alpha = \frac{n}{m}$ and fixed $\beta$. This is achieved in the following theorem.

*Theorem 12:* Given the static partitioning of the bipartite graph $G_\beta$, the limit normalized expected maximum matching size $\gamma_\beta = \lim_{n\to\infty} \frac{\mu(G_\beta)}{n}$ for $\beta \in (0,1)$ is given by: $\gamma_\beta = \frac{1}{\alpha} - \frac{\beta \cdot (1-\beta)}{\alpha^2} \cdot (t_1 + t_2 - t_1 \cdot t_2)$, where $t_1, t_2$ are provided by the following equations

$$\frac{\alpha}{1-\beta} \cdot e^{-\frac{\alpha}{\beta}} = t_1 \cdot e^{-t_2} \quad , \quad \frac{\alpha}{\beta} \cdot e^{-\frac{\alpha}{1-\beta}} = t_2 \cdot e^{-t_1} \quad (2)$$

and satisfy the condition $t_1 \cdot t_2 \leq 1$.

For $\beta \in \{0,1\}$, (namely, the trivial partitions), the limit normalized expected maximum matching size $\gamma_\beta$ is $\frac{1}{\alpha} - \frac{1}{\alpha} \cdot e^{-\alpha}$.

We deduce the following two corollaries.

*Corollary 13 (Asymptotic Equivalence):* Let $d = 2$. The limit normalized expected maximum matching size of $G_\beta$ with $\beta = 0.5$ is the same as the limit expected maximum matching size of $G$.

*Proof:* We substitute $\beta = 0.5$ in the expression from Theorem 12, and get $\frac{\alpha}{0.5} \cdot e^{-\frac{\alpha}{0.5}} = t_1 \cdot e^{-t_2}$, $\frac{\alpha}{0.5} \cdot e^{-\frac{\alpha}{0.5}} = t_2 \cdot e^{-t_1}$. One of the solutions of the above equations is $t_1 = t_2 = -W(-2\alpha e^{-2\alpha})$. In the proof of Theorem 4, we showed that $-W(-2\alpha e^{-2\alpha}) \leq 1$. Thus, $t_1 \cdot t_2 < 1$. By substituting this solution in the expression for $\gamma_\beta$ from Theorem 12, we get the exact expression as in Equation (1). ∎

*Corollary 14:* Let $d = 2$, $\alpha \leq \frac{1}{2}$, and fix a partition $\beta$. The limit normalized expected maximum matching size $\gamma_\beta = \lim_{n\to\infty} \frac{\mu(G_\beta)}{n}$ is 1 whenever $\frac{1-\sqrt{1-4\alpha^2}}{2} \leq \beta \leq \frac{1+\sqrt{1-4\alpha^2}}{2}$.

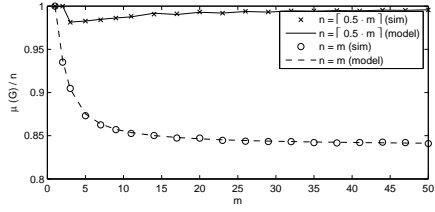

Fig. 2. Expected maximum matching size for various values of $n$ and $m$ (normalized by $n$)

### VII. SUPER-CUCKOO: BIPARTITE GRAPHS WITH $d > 2$

We are now interested in checking how powerful cuckoo hashing can be when we allow more than 2 hash functions per element. Of course, using more hash functions will result in an increase in implementation complexity, and therefore one goal of this study is to point out the tradeoff between efficiency and complexity.

In this section we briefly show how our method can be applied to find an upper bound on the expected maximum size matching where each left-side vertex has $d > 2$ choices. Formally, we are given two disjoint sets of vertices $L$ and $R$ of size $n$ and $m$, respectively, and a random bipartite graph $G^d = \langle L + R, E \rangle$, where each vertex $v \in L$ has $d$ outgoing edges whose destinations are chosen independently at random (with repetition) among all vertices in $R$. We obtain the following upper bound on the maximum matching size of the bipartite graph $G^d$.

*Theorem 15:* Let $b = \min\left\{n, \left\lfloor \frac{m-1}{d-1} \right\rfloor\right\}$ and $q = (d-1)\cdot s + 1$. Then, $\mu\left(G^d\right)$ is lower or equal to

$$\min\left\{n, m - \sum_{s=0}^{b}(q-s)\binom{n}{s}\binom{m}{q}\left(1 - \frac{q}{m}\right)^{d(n-s)}\left(\frac{q}{m}\right)^{ds}\frac{d^s \cdot q!}{q^{(d-1)\cdot s+2}}\right\}.$$

An evaluation of the upper bound and a comparison to the simulated expected matching size is presented in Section VIII-D.

### VIII. EVALUATION AND EXPERIMENTS

#### A. Expected Maximum Matching Size With $d = 2$

Figure 2 shows the expected maximum matching size normalized by $n$ for various values of $n$ and $m$. We show the expected maximum matching size both via our analytical model from Theorem 1 and via simulations. For each instance of $n$ and $m$, we randomized $m = 10{,}000$ bipartite graphs. The results fairly confirm that our model is accurate, and also show the convergence of the expected maximum matching size to its limit. A simple evaluation appears in the following example.

*Example 1:* In case $n = m = 2$ (and $d = 2$), the expected maximum matching size is $\mu(G) = \frac{15}{8} = 1.875$. This simple result can be justified as follows: In all cases the maximum matching size is 2, except for the two cases of maximum matching of size 1, where all 4 edges are connected to a specific vertex in $R$. Each such case occurs with probability $\left(\frac{1}{2}\right)^4$. Hence, $\mu(G) = 2 - \frac{1}{16} - \frac{1}{16} = \frac{15}{8}$.

Figure 3 shows the expected maximum matching size normalized by $n$ as found in Theorem 4, for various values of load $\alpha$, both via our analytical model and via simulations. The simulations were performed using $m = 1000$ and $n = \alpha \cdot m$. For each value of $\alpha$, we randomized 100 bipartite graphs. The results fairly confirm that our model is accurate.

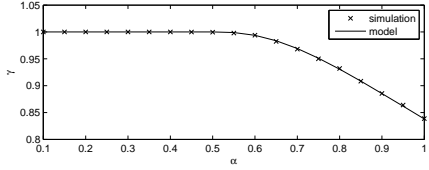

Fig. 3. Limit expected maximum matching size for various values of load $\alpha$, normalized by $n$

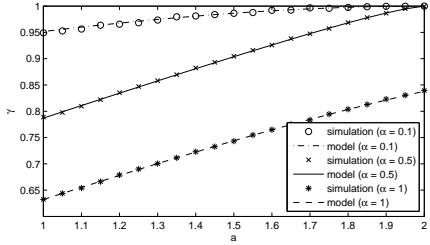

Fig. 4. Limit expected maximum matching size for various values of $\alpha$ and $a$, normalized by $n$

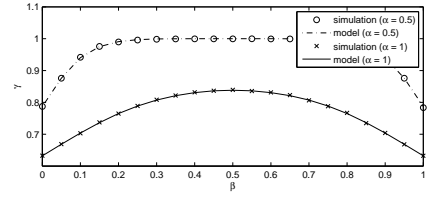

Fig. 5. Limit expected maximum matching size for various values of $\beta$ and $\alpha$, normalized by $n$

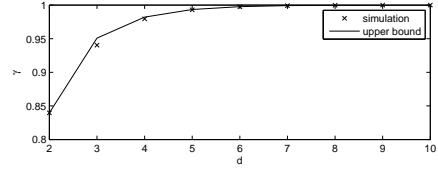

Fig. 6. Upper bound on the normalized expected maximum matching size for $\alpha = 1$ as a function of $d$

We conclude by the following simple example:

*Example 2:* In case $\alpha = 1$, that is $n = m$, the normalized limit expected maximum matching size is

$$\gamma = 1 + \frac{1}{2} \cdot W\left(-2 \cdot e^{-2}\right) + \frac{1}{4} W^2\left(-2 \cdot e^{-2}\right) \approx 0.8381.$$

### B. Expected Maximum Matching Size With $d_v \leq 2$

Figure 4 shows the normalized limit expected maximum matching size, for various values of load $\alpha$ and average number of choices $a$, both via our analytical model (from Theorem 4) as well as via simulations. The simulations were performed using $m = 1000$ and $n = \alpha \cdot m$, where for each instance of the simulation we randomized 100 bipartite graphs. The results fairly confirm that our model is accurate.

### C. Expected Maximum Matching Size With Static Partition

Figure 5 shows the limit expected maximum matching size normalized by $n$, for various values of load $\alpha$ and partition $\beta$, both via our analytical model (from Theorem 12) and via simulations. The simulations were performed using $m = 1000$ and $n = \alpha \cdot m$. For each pair of values of $\alpha$ and $\beta$, we randomized 100 bipartite graphs. The results fairly confirm that our model is accurate.

As expected, the limit expected maximum matching size is symmetric around $\beta = 0.5$. In case $\alpha = 0.5$ and $\beta < 0.5$, while it seems that the normalized limit expected maximum matching size is 1, it is not the case. For instance, in case $\alpha = 0.5$ and $\beta = 0.45$, we get that $1 - \gamma_\beta \approx 1.675 \cdot 10^{-7}$. However, there are cases where imbalance in the partition sizes does not reduce $\gamma_\beta$, as shown for instance in Corollary 14.

### D. Expected Maximum Matching Size With $d > 2$

We evaluate the upper bound found for the expected matching size (Theorem 15). Figure 6 shows our upper bound as well as simulation results for various values of the number of choices $d$. We took $n = m = 100$, while for each instance of $d$, we randomized $10^5$ bipartite graphs. In the case of $d = 2$, our upper bound matches the exact expression found in Theorem 1 and thus matches the simulation results. In addition, we can compare simulation results for higher values of $d$ with our bounds. For instance, in the case of $d = 3$ the normalized expected maximum matching size via the simulation is $0.9402$, while our upper bound is $0.9508$. In case $d = 4$, we get a simulation value of $0.9795$, while the corresponding upper bound is $0.9820$.

### E. Trace-Driven Experiments

We have also conducted experiments using real-life traces recorded on a single direction of an OC192 backbone link [41], where packets are hashed using a real 64-bit mix function [42]. Our goal is two-folded. First, we would like to verify that our analysis agrees with results of real-life traces. And second, we want to verify that the distribution of overflow list size is highly concentrated around its mean, as stated in Theorem 2.

We took $m = 10,000$, and set a number of elements $n$ as corresponding to various values of load $\alpha$. We repeated each experiment 100 times. Fig. 7 shows that the results of our experiments are very close to our model. Furthermore, it also shows the minimum and the maximum overflow list size as a function of the load $\alpha$, thus, introducing a *confidence interval* of 98% for the case where $m = 10,000$. Note that, as reflected in Thoerem 2, if we increase $m$ (and set $n$ accordingly) then the confidence interval narrows down.

## IX. CONCLUSION

In this work, we analyzed the performance of cuckoo hashing with a stash for loads above 0.5. We first provided an exact expression for the expected maximum matching size of a random bipartite graph with each left-side vertex picking $d = 2$ right-side vertices. Then, we deduced asymptotic results as the memory size goes to infinity, and showed a connection to the Lambert-$W$ function.

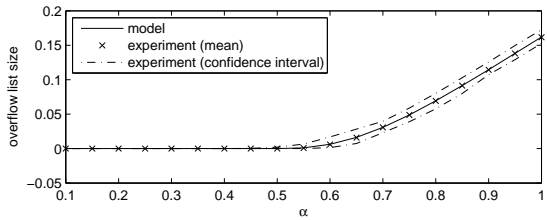

Fig. 7. An experiment using real-life traces

Both these results directly apply as exact results for the average number of inserted elements using cuckoo hashing. They also help us size the stash needed in the algorithm. In addition, they serve as an upper bound for any alternative hashing algorithm.

We also discussed alternative cases, in which cuckoo either uses a lower memory bandwidth to gain power, or uses a higher memory bandwidth to gain in efficiency, as well as a case in which memory is partitioned and can be implemented using two single-ported memories. Finally, we evaluated our results on Internet backbone traces.

As future work, our goal is to implement the algorithm in FPGAs, and evaluate its performance according to the measures accepted in the switch industry (e.g., mean time to failure vs. power utilization resulting from the CAM stash).

ACKNOWLEDGMENT

This work was partly supported by the European Research Council Starting Grants n°210389 and n°259085, the Alon Fellowship, the ATS-WD Career Development Chair, and the Loewengart Research Fund.

## APPENDIX A
## OMITTED PROOFS

### A. Proof of Lemma 1

The proof follows by induction on $s$. For $s = 1$, there are 2 edges in the graph and therefore every graph with $q \geq 3$ is not connected. Assume that the claim holds up until $s = s'$, we next prove that it holds for any bipartite graph $H'$ such that $|L_{H'}| = s' + 1$ and $|R_{H'}| \geq s' + 3$. Assume towards a contradiction that there is a graph $H'$ that is connected. We first show that there is a vertex in $R_{H'}$ with a degree 1: This follows from the fact that the average right-side degree is $\frac{2(s'+1)}{s'+3} < 2$, implying that there is at least one vertex with degree strictly less than 2; since the graph is connected, there are no right-side vertices with degree 0. Let $v_r$ be such a vertex and let $v_\ell \in L_{H'}$ be the (only) left-side vertex to which it is connected. By the induction hypothesis, the graph induced by $L_{H'} \setminus \{v_\ell\}$ and $R_{H'} \setminus \{v_r\}$ is not connected, implying it has at least two connected components. In $H'$, $v_\ell$ is connected to $v_r$ and since its degree is 2 it can be connected only to one of these components. This implies that $H'$ is also not connected, and the claim follows. ∎

### B. Proof of Lemma 2

We first consider the case where $s = q$. For $S \subseteq L_H$, let $d(S) \subseteq R_H$ be the set of vertices that are adjacent to any vertex in $S$. Hall's Theorem [43] implies that to prove that $\mu(H) = q$ (namely, there is a perfect matching in $H$) it suffices to prove that for every $S \subseteq L_H$, $|S| \leq |d(S)|$. Assume towards a contradiction that there is a subset $S \subseteq L_H$ such that $|S| > |d(S)|$, and denote $|d(S)|$ by $b$. Furthermore, consider the bipartite graph $\hat{H} = \left\langle \hat{L}_H + \hat{R}_H, \hat{E}_H \right\rangle$, in which $\hat{L}_H = L_H \setminus S$, $\hat{R}_H = R_H \cup \{\hat{v}_R\} \setminus d(S)$ (where $\hat{v}_R$ is a newly-introduced vertex) and any edge in $E(H)$ of the form $(v_\ell, v_r)$ such that $v_\ell \in L_H \setminus S$ and $v_r \in d(S)$ is replaced with the edge $(v_\ell, \hat{v}_R)$ in $\hat{E}_H$. Notice that since $H$ is connected, $\hat{H}$ must be connected as well. Recall that $|S| > b$, thus $\left|\hat{L}_H\right| = |L_H \setminus S| \leq s - b - 1$, while $\left|\hat{R}_H\right| = |R_H \cup \{\hat{v}_R\} \setminus d(S)| = |R_H| - |d(S)| + 1 = s - b + 1$. This contradicts Lemma 1, implying that for every $S \subseteq L_H$, $|S| \leq |d(S)|$ and by Hall's Theorem $\mu(H) = q$.

For $s > q$, trivially $\mu(H) \leq q$. Therefore, it suffices to show that there exists a subset $S \subseteq L_H$ of size $q$, such that the corresponding bipartite subgraph is connected (and hence has a perfect matching of size $q$). We construct $S$ in $q$ iterations such that at the end of iteration $n$ we end up with some subsets $S_n \subseteq L_H$ and $Q_n \subseteq R_H$ of the same size $n$, whose corresponding subgraph is connected. We start by $n = 1$ and pick some vertex $v_R \in R_H$ and one of its adjacent vertices $v_L \in L_H$. Assuming that at the end of iteration $n$, sets $S_n$ and $Q_n$ were chosen (and their corresponding graph is connected), we next construct $S_{n+1}$ and $Q_{n+1}$. Let $v_1$ be an arbitrary vertex in $S_n$ and let $v_2$ be an arbitrary vertex in $L_H \cup S_n$ (such a vertex always exists since $s > q > n$). Similarly, let $v'_1$ be an arbitrary vertex in $Q_n$ and let $v'_2$ be an arbitrary vertex in $R_H \cup Q_n$. Since $H$ is connected there is a path between $v_1$ and $v_2$, and let $v$ be the first vertex along this path that is not in $S_n$. Similarly, $v'$ is the first vertex along the path between $v'_1$ and $v'_2$ that is not in $Q_n$. We differentiate between three cases: *(i)* $v$ is adjacent to $Q_n$ and $v'$ is to $S_n$. In this case $S_{n+1} = S_n \cup \{v\}$ and $Q_{n+1} = Q_n \cup \{v'\}$ and the corresponding subgraph is connected; *(ii)* $v$ is not adjacent to a $Q_n$. Let $w$ be the vertex before $v$ in the path between $v_1$ and $v_2$, and let $w'$ be the vertex before $w$ in the path. Note that $w' \in S_n$ by the choice of $v$, and that $w \notin Q_n$ (otherwise $v$ is adjacent to a $Q_n$). Thus, for $S_{n+1} = S_n \cup \{v\}$ and $Q_{n+1} = Q_n \cup \{w\}$, the corresponding subgraph is connected; *(iii)* $v'$ is not adjacent to a $S_n$. The claim holds similarly to case (ii) by looking at the path between $v'_1$ and $v'_2$. We continue this construction for $q$ iterations, resulting in two subsets $S_q \subseteq L_H$ and $Q_q \subseteq R_H$ of size $q$ each, whose corresponding subgraph is connected. ∎

### C. Proof of Lemma 3

Since each vertex in $L_H$ has a degree of two, the sum of the degrees of all the vertices in $R_H$ is $2s = 2q - 2$. Therefore, there must be at least one vertex $v_r \in R_H$ with degree 1 (there cannot be a vertex with degree 0 since $H$ is connected). Let $v_L \in L_H$ be the (only) vertex that is connected to $v_R$ and $\hat{v}_R \in R_H$ be the other vertex that is connected to $v_L$. Also consider the bipartite graph $\hat{H} = \left\langle \hat{L}_H + \hat{R}_H, \hat{E}_H \right\rangle$ that is given by removing $v_R$ from $H$ and adding a new edge $(v_L, \hat{v}_R)$. By the construction of $\hat{H}$, the degree of each vertex in $\hat{L}_H$ is exactly 2. Moreover, since $H$ is connected, $\hat{H}$ is also connected. Hence, Lemma 2 implies that there is a matching of size $s$ in $\hat{H}$. By the construction of $\hat{H}$, this is also a matching in graph $H$. ∎

### D. Proof of Lemma 4

First, if $H$ is a tree then it is connected by definition. To show the other direction, we assume towards a contradiction that $H$ is a connected graph with cycles; let $C$ be a cycle in $H$, and consider an edge $e = (v_L, v_R)$ that resides at cycle $C$ (where $v_L \in L_H$ and $v_R \in R_H$). We build the bipartite graph $\hat{H} = \left\langle \hat{L}_H + \hat{R}_H, \hat{E}_H \right\rangle$, such that $\hat{L}_H = L_H$, $\hat{R}_H = R_H \cup \{\hat{v}_R\}$, where $\hat{v}_R$ is a newly-introduced vertex, and $\hat{E}_H = E_H \setminus \{e\} \cup \{\hat{e}\}$, where $\hat{e} = (v_L, \hat{v}_R)$. Intuitively, we replace one of the edges in the cycle to reach for a newly-introduced vertex, and by that we increase the size of the connected component. Notice that $\hat{H}$ is connected and all vertices in $\hat{L}_H$ have a degree of 2. But, $\left|\hat{L}_H\right| < \left|\hat{R}_H\right| - 1$, thus contradicting Lemma 1 and the claim follows. ∎

### E. Proof of Lemma 5

We count the connected bipartite graphs with two disjoint sets $L_H$ and $R_H$. By Lemma 4, we have to count the number of trees over the set $L_H \cup R_H$, where edges must be of the form $(v_L, v_R)$, such that $v_L \in L_H$ and $v_R \in R_H$. We build (and count) the set as follows: The number of trees over the set $R_H$ is $(s+1)^{s-1}$. For each such tree instance, we put a new vertex (originally from $L_H$) between each pair of adjacent vertices. There are $s!$ possibilities to do so. ∎

### F. An Alternative Proof of Theorem 4

Considering the random graph with $m$ vertices and $n$ edges such that a vertex $m_1$ is connected to vertex $m_2$ if and only if there exists an element that hashes into $m_1$ and $m_2$. This random graph is called the cuckoo graph [9]. Neglecting the $O(1)$ loops, this graph is equivalent to the Erdös-Renyi random graph $G_{m,n}$ that assigns equal probability to all graphs with exactly $n$ edges (and $m$ vertices)

A matching in $G_{m,n}$ corresponds to directing some of the edges in the random graph such that the in-degree is at most 1. For each connected component $C$ in $G_{m,n}$, if $C$ is a tree we can direct all edges, while in all other cases we can direct as much edges as the number of vertices.

The number of such edges and vertices can be found in [34], [37], yielding the exact same result.

### G. Proof of Lemma 7

The proof is identical to the proof of Lemma 4 with two modifications. First, instead of initially counting the number of trees over the set $R_H$, we count the number of parity trees [44] over the disjoint sets $R_{H_u}$ and $R_{H_d}$. By [44] we are given that the number of parity trees is $i^{j-1} \cdot j^{i-1}$. Second, we do not have to color the edges because of the partition. ∎

### H. Proof of Theorem 12

As in the proof of Theorem 4, we compute the limit of $\frac{\mu(G)}{n}$ as $n \to \infty$. We consider the case where $\alpha = \frac{n}{m}$ and $0 \le \beta \le 1$ are fixed. So $\gamma_\beta = \lim_{n \to \infty} \frac{\mu(G_\beta)}{n}$, that is,

$$\gamma_\beta = \lim_{n \to \infty} \frac{1}{n} \cdot \left( m - \sum_{s=0}^{n} \binom{n}{s} \cdot \sum_{i=b_1}^{b_2} \binom{\beta \cdot m}{i} \binom{(1-\beta) \cdot m}{s+1-i} \cdot \left(1 - \frac{i}{\beta \cdot m}\right)^{n-s} \cdot \left(1 - \frac{s+1-i}{(1-\beta) \cdot m}\right)^{n-s} \cdot \left(\frac{i}{\beta \cdot m}\right)^s \cdot \left(\frac{s+1-i}{(1-\beta) \cdot m}\right)^s \cdot P_{i, s+1-i} \right)$$

By substituting the expression for $P_{i,s+1-i}$ from Theorem 11, and moving $\binom{n}{s}$ inside the second summation, we get:

$$\gamma_\beta = \lim_{n \to \infty} \left( \frac{1}{\alpha} - \frac{1}{n} \sum_{s=0}^{n} \sum_{i=0}^{s+1} \binom{n}{s} \binom{\beta m}{i} \binom{(1-\beta)m}{s+1-i} \left(1 - \frac{i}{\beta m}\right)^{n-s} \right. $$
$$\left(1 - \frac{s+1-i}{(1-\beta) \cdot m}\right)^{n-s} \cdot \left(\frac{i}{\beta \cdot m}\right)^s \cdot \left(\frac{s+1-i}{(1-\beta) \cdot m}\right)^s \cdot $$
$$\left. \frac{i^{(s+1-i)-1} \cdot (s+1-i)^{i-1} \cdot (i+(s+1-i)-1)!}{(i \cdot (s+1-i))^{i+(s+1-i)-1}} \right)$$

By substituting $\alpha = \frac{n}{m}$, we get:

$$\gamma_\beta = \lim_{n \to \infty} \left( \frac{1}{\alpha} - \frac{1}{n} \sum_{s=0}^{n} \sum_{i=0}^{s+1} \binom{n}{s} \binom{\frac{\beta}{\alpha}n}{i} \binom{\frac{1-\beta}{\alpha}n}{s+1-i} \left(1 - \frac{i}{\frac{\beta}{\alpha}n}\right)^{n-s} \right. $$
$$\left(1 - \frac{s+1-i}{\frac{1-\beta}{\alpha} \cdot n}\right)^{n-s} \cdot \left(\frac{i}{\frac{\beta}{\alpha} \cdot n}\right)^s \cdot \left(\frac{s+1-i}{\frac{1-\beta}{\alpha} \cdot n}\right)^s \cdot $$
$$\left. \frac{i^{(s+1-i)-1} \cdot (s+1-i)^{i-1} \cdot (i+(s+1-i)-1)!}{(i \cdot (s+1-i))^{i+(s+1-i)-1}} \right)$$

As in the proof of Theorems 4 and 8, using the monotone convergence theorem [36], we can put the limit inside the sum. By further simplifying the above expression with similar consideration to the proofs of Theorems 4 and 8, we get eventually:

$$\gamma_\beta = \frac{1}{\alpha} - \frac{\beta \cdot (1-\beta)}{\alpha^2} \sum_{s=0}^{\infty} \sum_{i=0}^{s+1} \frac{i^{(s+1-i)-1} \cdot (s+1-i)^{i-1}}{i! \cdot (s+1-i)!} \cdot $$
$$\left(\frac{\alpha}{\beta} \cdot e^{-\frac{\alpha}{1-\beta}}\right)^{s+1-i} \cdot \left(\frac{\alpha}{1-\beta} \cdot e^{-\frac{\alpha}{\beta}}\right)^i$$

We switch the order of summation and get that $i \in \{0, 1, \ldots\}$ and $s$ goes from $\max\{0, i-1\}$ to $\infty$. We also substitute $j = s+1-i$ (or $s = i+j-1$). Thus,

$$\gamma_\beta = \frac{1}{\alpha} - \frac{\beta \cdot (1-\beta)}{\alpha^2} \sum_{i=0}^{\infty} \sum_{j=\max\{0,i-1\}}^{\infty} \frac{i^{j-1} \cdot j^{i-1}}{i! \cdot j!} \cdot \quad (3)$$
$$\left(\frac{\alpha}{1-\beta} \cdot e^{-\frac{\alpha}{\beta}}\right)^i \cdot \left(\frac{\alpha}{\beta} \cdot e^{-\frac{\alpha}{1-\beta}}\right)^j$$

Let $T(x,y) = \sum_{j+i \ge 1} \frac{i^{j-1} \cdot j^{i-1}}{i! \cdot j!} \cdot x^i \cdot y^j$. This expression has been previously found [7] to be the multivariate formal power series about the point $(x_0, y_0) = (0, 0)$ of $t(x, y) = t_1(x, y) + t_2(x, y) - t_1(x, y) \cdot t_2(x, y)$ where $t_1(x, y)$ and $t_2(x, y)$ are given by the following implicit multivariate functions:

$$x = t_1(x,y) \cdot e^{-t_2(x,y)} \quad , \quad y = t_2(x,y) \cdot e^{-t_1(x,y)} \quad (4)$$

However, the mentioned range of convergence in [7] is insufficient for our case. (Note also that in [7] the sums should be over $i + j \ge 1$ and not over $i, j \ge 0$.)

Since we compute the limit normalized expected maximum matching, then the expression for $\gamma_\beta$ in Equation (3) is bounded from below by 0, thus, by Equation (3) the double summation is bounded from above by a constant. On the other hand, all terms in the summation in Equation (3) are positive. Then, if we look at the partial-sum series (by defining an arbitrary order), we get an increasing series which is bounded. Thus, by the monotone convergence theorem the double series converges for any values $x$ and $y$ satisfying $x = \frac{\alpha}{1-\beta} \cdot e^{-\frac{\alpha}{\beta}}$ and $y = \frac{\alpha}{\beta} \cdot e^{-\frac{\alpha}{1-\beta}}$.

However, the multivariate functions in Equation (4) have multiple branches (as the Lambert-$W$ function does [35]), that is, for a given $x$ and $y$ there is more than one solution. We aim to find this branch in terms of $t_1$ and $t_2$. We use the implicit function theorem to find the derivatives singularities. The Jacobian is given by

$$J = \begin{pmatrix} e^{-t_2(x,y)} & -t_1(x,y) \cdot e^{-t_2(x,y)} \\ -t_2(x,y) \cdot e^{-t_1(x,y)} & e^{-t_1(x,y)} \end{pmatrix},$$

and it is invertible wherever $|J| \neq 0$. Thus, there is a derivative singularity in case $t_1(x,y) \cdot t_2(x,y) = 1$, which is the only solution. Therefore, as the given formal power series in Equation (3) is about the point $(x_0, y_0) = (0, 0)$ (which corresponds to $\alpha = 0$), where $t_1 = t_2 = 0$, it converges to the branch where $t_1(x, y) \cdot t_2(x, y) \leq 1$ (note that both $t_1(x, y)$ and $t_2(x, y)$ are always positive). ∎

### I. Proof of Corollary 14

One of the solutions to Equation (2) is given by: $t_1 = \frac{\alpha}{1-\beta}$, $t_2 = \frac{\alpha}{\beta}$. By substituting $t_1$ and $t_2$ in the expression for $\gamma_\beta$ from Theorem 12, we get that the limit normalized expected maximum matching size is 1. We also have to verify that $t_1 \cdot t_2 \leq 1$. Since $\frac{\alpha}{1-\beta}$ and $\frac{\alpha}{\beta}$ are both positive, we are left with $\frac{\alpha}{1-\beta} \cdot \frac{\alpha}{\beta} < 1$. By solving the quadratic inequality, we get the claimed condition. Note that for $\alpha = 1/2$ the range reduces to $\beta = 1/2$. ∎

### J. Proof of Lemma 6

Assume on the contrary that $H$ is connected but that there is (at least) a single vertex $v_L \in L_H$ with degree 1. Consider the bipartite graph $\hat{H} = \left\langle \hat{L}_H + \hat{R}_H, \hat{E}_H \right\rangle$, that is given by removing the vertex $v_L$ (and its connected edge) from $H$. By the construction of $\hat{H}$, we get that $\hat{H}$ is connected, but $\left|\hat{L}_H\right| + 1 < \left|\hat{R}_H\right|$, which contradicts Lemma 1. ∎

### K. Proof of Theorem 6

As in the proof of Theorem 1, our proof is based on counting the expected number of vertices in $L$ that are not in some specific maximum matching $M$ of $G$, based on the decomposition of $G$ into its connected components. The proof is almost identical, with the modification that, due to Lemma 6, we only take into account the $d_2$ vertices that have a degree of 2 (instead of all $n$ vertices in the proof of Theorem 1).

Thus, the expected number of connected components in $G$ with $s$ elements in $L$ and $s+1$ in $R$ is given by:

$$\binom{d_2}{s}\binom{m}{s+1} \cdot \left(1 - \frac{s+1}{m}\right)^{2(d_2-s)+d_1} \cdot \left(\frac{s+1}{m}\right)^{2s} \cdot P_s,$$

where the above expression consists of the same considerations as in the proof of Theorem 1. Finally, as before, adding the expressions for all possible $s$'s and subtracting the sum from $m$ yields the claimed result. ∎

### L. Proof of Theorem 7

The number of vertices in $L$ with degree 2 follows a Binomial distribution with $n$ experiments and a probability of success $p$. In Theorem 6 we found the expected maximum matching size of each such instance. Thus, by the law of total expectation, the claimed result is given by computing the weighted average, where we compute $a$ by the equations $d_1 + d_2 = n$ and $d_1 + 2 \cdot d_2 = a \cdot n$. ∎

### M. Proof of Theorem 8

We compute the limit of $\frac{\mu(G_a)}{n}$ as $n \to \infty$. We consider the case where $\alpha = \frac{n}{m}$ and $a = \frac{d_1 + 2 \cdot d_2}{n} > 1$ are fixed. So $\gamma_a = \lim_{n \to \infty} \frac{\mu(G_a)}{n}$, that is,

$$\gamma_a = \lim_{n \to \infty} \frac{1}{n} \left( m - \sum_{s=0}^{b} \binom{d_2}{s} \binom{m}{s+1} \left(1 - \frac{s+1}{m}\right)^{2(d_2-s)+d_1} \cdot \left(\frac{s+1}{m}\right)^{2s} \cdot P_s \right).$$

Given that $a = \frac{d_1 + 2 \cdot d_2}{n}$ and $n = d_1 + d_2$, we find that $d_2 = (a-1) \cdot n$ and $d_1 = (2-a) \cdot n$. Similarly to the proof of Theorem 4, we first have to find that each term in the summation is an increasing function with respect to $n$. We discover that $\left(1 - \frac{s+1}{m}\right)^{2(d_2-s)+d_1} = \left(1 - \frac{s+1}{m}\right)^{a \cdot n - s}$ is an increasing function (using differentiation), and also find that $\frac{1}{n} \cdot \binom{(a-1) \cdot n}{s}\binom{m}{s+1} \cdot \left(\frac{s+1}{m}\right)^{2s}$ is an increasing function as previously. Consequentially, each term in the sum is an increasing function and, by the monotone convergence theorem [36], we can put the limit inside the sum. By further simplifying the above expression as in the proof of Theorem 4 we eventually get:

$$\gamma_a = \frac{1}{\alpha} - \frac{1}{2\alpha^2 \cdot (a-1)} \cdot \sum_{j=1}^{\infty} \frac{(-j)^{j-2}}{j!} \cdot \left(-\alpha \cdot 2 \cdot (a-1) \cdot e^{-a\alpha}\right)^j$$

Let $T(x) = \sum_{j=1}^{\infty} \frac{(-j)^{j-2}}{j!} \cdot x^j$ be a Taylor expansion, where by substituting $x = -\alpha \cdot 2 \cdot (a-1) \cdot e^{-a\alpha}$ we get the above expression. Similarly to the proof of Theorem 4, we get that

$$T(x) = -W(x) - \frac{1}{2}W^2(x),$$

with convergence within $|x| \leq e^{-1}$ [35].

Since the function $f(\alpha) = -\alpha \cdot 2 \cdot (a-1) \cdot e^{-a\alpha}$ gets its minimum at $\alpha = a^{-1}$, where it equals $-\frac{2(a-1)}{a}e^{-1}$, and $\left|-\frac{2(a-1)}{a}e^{-1}\right| \leq e^{-1}$ for all $a \in [1, 2]$, then for all $\alpha$ we can substitute $x = -\alpha \cdot 2 \cdot (a-1) \cdot e^{-a\alpha}$. Hence, it is within the radius of convergence of $T(x)$.

Finally, for the case where $a = 1$, then $d_2 = 0$ and $d_1 = n$. Therefore, the expression for the expected maximum matching size is reduced to $m - \left(m \cdot \left(1 - \frac{1}{m}\right)^n\right)$. Thus,

$$\gamma_a = \lim_{n \to \infty} \frac{\mu(G_a)}{n} = \lim_{n \to \infty} \frac{1}{n} \cdot \left(m - \left(m \cdot \left(1 - \frac{1}{m}\right)^n\right)\right)$$
$$= \frac{1}{\alpha} - \frac{1}{\alpha} \cdot e^{-\alpha}.$$
∎

### N. Proof of Corollary 9

We show that $\gamma_a$ is strictly monotonically increasing, thus $\gamma_a < 1$ for $1 \leq a < 2$, since $\gamma_a = 1$ for $a = 2$. This is shown by differentiating $\gamma_a$ with respect to $a$:

$$\frac{d\gamma_a}{da} = -\frac{1}{4\alpha^2 (a-1)^2} \cdot \left(W\left(-2\alpha(a-1) \cdot e^{-a\alpha}\right) + 2\alpha(a-1) \cdot W\left(-2\alpha(a-1) \cdot e^{-a\alpha}\right)\right)$$

Both the first factor $-\frac{1}{4\alpha^2(a-1)^2}$ and the third factor $W\left(-2\alpha(a-1) \cdot e^{-a\alpha}\right)$ are negative. Thus, if the second factor is positive then $\frac{d\gamma_a}{da}$ is an increasing function with respect to $a \in [1, 2)$.

If $\alpha > 0.5$, then $2\alpha(a-1) > 1$, and since $W(x)$ is minimized for $x = -\frac{1}{e}$ where it equals $-1$, the second factor is positive. On the other hand, consider that $\alpha \leq 0.5$. Since $W\left(-2\alpha(a-1) \cdot e^{-2\alpha(a-1)}\right) = -2(a-1)\alpha$ and $W(x)$ is an increasing function, then we have to show that $-2\alpha(a-1) \cdot e^{-2\alpha(a-1)} < -2\alpha(a-1) \cdot e^{-a\alpha}$, that is, $-2\alpha(a-1) > -a\alpha$. The last inequality can easily be shown for $1 \leq a < 2$. ∎

## O. Proof of Theorem 10

We compute the limit of $\frac{\mu(G_p)}{n}$ as $n \to \infty$.

$$\gamma_p = \lim_{n \to \infty} \frac{\mu(G_p)}{n}$$
$$= \lim_{n \to \infty} \frac{1}{n} \sum_{d_2=0}^{n} \binom{n}{d_2} \cdot p^{d_2} \cdot (1-p)^{n-d_2} \cdot \mu\left(G_{a=1+\frac{d_2}{n}}\right)$$

Let $X \sim \text{Bin}(n, p)$ be the random variable counting the number of vertices in $L$ that choose 2 vertices in $R$. By summing over three disjoint ranges of possible values for $d_2$, we get

$$\gamma_p = \lim_{n \to \infty} \sum_{d_2=0}^{\lfloor np - n^{\frac{3}{4}}\rfloor} \Pr\{X = d_2\} \cdot \frac{1}{n} \cdot \mu\left(G_{a=1+\frac{d_2}{n}}\right) +$$
$$\lim_{n \to \infty} \sum_{d_2=\lfloor np - n^{\frac{3}{4}}\rfloor+1}^{\lfloor np + n^{\frac{3}{4}}\rfloor-1} \Pr\{X = d_2\} \cdot \frac{1}{n} \cdot \mu\left(G_{a=1+\frac{d_2}{n}}\right) +$$
$$\lim_{n \to \infty} \sum_{d_2=\lfloor np - n^{\frac{3}{4}}\rfloor}^{n} \Pr\{X = d_2\} \cdot \frac{1}{n} \cdot \mu\left(G_{a=1+\frac{d_2}{n}}\right)$$

By Chebyshev's inequality we get that $\Pr\left\{|X - np| > n^{\frac{1}{4}}\sqrt{np(1-p)}\right\} \leq \frac{1}{n^{\frac{1}{4}}}$. Since $p(1-p) \leq 1$, we get that $\Pr\left\{|X - np| > n^{\frac{3}{4}}\right\} \leq \frac{1}{n^{\frac{1}{4}}}$. By the fact that $\frac{1}{n} \cdot \mu\left(G_{a=1+\frac{d_2}{n}}\right) \leq 1$, we find that the first and the third limits go to zero.

Since the function $\mu(G_a)$ is increasing with respect to $a$ (this can be shown by a simple combinatorial argument), we get the following lower bound:

$$\gamma_p = \lim_{n \to \infty} \sum_{d_2=\lfloor np-n^{\frac{3}{4}}\rfloor+1}^{\lfloor np+n^{\frac{3}{4}}\rfloor-1} \Pr\{X = d_2\} \cdot \frac{1}{n} \cdot \mu\left(G_{a=1+\frac{d_2}{n}}\right)$$
$$\geq \lim_{n \to \infty} \left(1 - \frac{1}{n^{\frac{1}{4}}}\right) \cdot \frac{1}{n} \cdot \mu\left(G_{a=1+\frac{\lfloor np-n^{\frac{3}{4}}\rfloor+1}{n}}\right)$$

as well as the following upper bound:

$$\gamma_p = \lim_{n \to \infty} \sum_{d_2=\lfloor np-n^{\frac{3}{4}}\rfloor+1}^{\lfloor np+n^{\frac{3}{4}}\rfloor-1} \Pr\{X = d_2\} \cdot \frac{1}{n} \cdot \mu\left(G_{a=1+\frac{d_2}{n}}\right)$$
$$\leq \lim_{n \to \infty} 1 \cdot \frac{1}{n} \cdot \mu\left(G_{a=1+\frac{\lfloor np+n^{\frac{3}{4}}\rfloor-1}{n}}\right).$$

By the squeeze theorem, we get the claimed result. ∎

## P. Proof of Theorem 15

We first establish a few lemmas before proving the result. As before, we start by considering a deterministic bipartite graph $H = \langle L_H + R_H, E_H \rangle$ with degree $d$ of each vertex in $L_H$, where $|L_H| = s$ and $|R_H| = q$.

*Lemma 8:* If $(d-1) \cdot s \leq q - 2$, then $H$ is not connected.

*Proof:* As in the proof of Lemma 1, the proof follows by induction on $s$. For $s = 1$, there are $d$ edges in the graph and therefore every graph with $q \geq d+1$ is not connected. Assuming that the claim holds up until $s = s'$, we next prove that it holds for any bipartite graph $H'$ such that $|L_{H'}| = s' + 1$ and $|R_{H'}| \geq (d-1) \cdot (s'+1) + 2$. Assume towards a contradiction that there is a graph $H'$ which is connected.

We first show that there are $d-1$ vertices $v_{r_1}, v_{r_2}, \ldots, v_{r_{d-1}}$ in $R_{H'}$, all of a degree 1 such that they are connected to the same vertex $v_\ell \in R_{H'}$: The sum of right-side vertex degree is $d \cdot (s'+1)$. Also, since the graph is connected there are no right-side vertices with degree 0. This implies that there are at least $(d-2) \cdot (s'+1) + 2$ vertices of degree 1, thus there exists a vertex $v_\ell \in R_{H'}$ as claimed.

By the induction hypothesis, the graph induced by $L_{H'} \setminus \{v_\ell\}$ and $R_{H'} \setminus \{v_{r_1}, v_{r_2}, \ldots, v_{r_{d-1}}\}$ is not connected, which implies that it has at least two connected components. In $H'$, $v_\ell$ is connected to all vertices $v_{r_1}, v_{r_2}, \ldots, v_{r_{d-1}}$. Since its degree is $d$ it can be connected only to one of these components. This implies that $H'$ is not connected as well, and the claim follows. ∎

*Lemma 9:* If $H$ is connected and $(d-1) \cdot s = q - 1$ then $\mu(H) = s$.

*Proof:* Assume towards a contradiction that $\mu(H) < s$, and consider some maximum matching $M$. Let $v_\ell \in L_H$ be a vertex that is not in the maximum matching $M$, and $v_{r_1}, v_{r_2}, \ldots, v_{r_{d-1}}$ be the vertices in $R$ (which are not necessarily distinct) that are connected to $v_\ell$. All vertices $v_{r_1}, v_{r_2}, \ldots, v_{r_{d-1}}$ are connected also to another vertex in $L_H$, otherwise $v_\ell$ was in the maximum matching $M$.

Consider the bipartite graph $\hat{H} = \langle \hat{L}_H + \hat{R}_H, \hat{E}_H \rangle$, which is given by removing $v_\ell$ from $H$. Since the right-side vertices $v_{r_1}, v_{r_2}, \ldots, v_{r_{d-1}}$ are also connected to the other left-side vertices (except $v_\ell$), the bipartite graph $\hat{H}$ is connected. However, we get that $\left|\hat{L}_H\right| = s-1$ and $\left|\hat{R}_H\right| = (d-1) \cdot s + 1$, which contradicts with Lemma 8. ∎

We note that in contrast to Lemma 2, the corresponding proposition is not true for $d > 2$; that is, if $H$ is connected and $s \leq q$, then the maximum matching size is not necessarily $s$. As a counter example, consider the case where $d = 3$ and $s = q = 3$, where two left-side vertices choose the same single right-side vertex (using all their 3 choices), and the other left-side vertex chooses all 3 right-side vertices. The resulting bipartite graph is clearly connected, but the maximum matching size is only 2 (only one of the first two left-vertices can be in the matching).

*Lemma 10:* If $(d-1) \cdot s = q - 1$ then $H$ is connected if and only if it is a tree.

*Proof:* The proof consists of the exact same construction $\hat{H}$ as in the proof of Lemma 4, where we eventually get a contradiction with Lemma 8. ∎

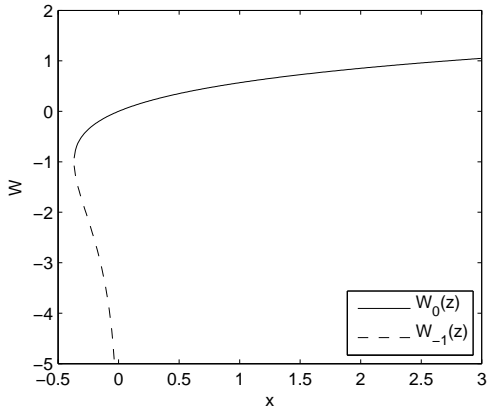

Fig. 8. the Lambert-$W$ function

*Lemma 11:* The number $T_s^d$ of connected bipartite graphs $H$ whose $|L_H| = s$ and $|R_H| = 2(d-1) \cdot s + 1$ is $T_s^d = \frac{((d-1) \cdot s+1)!}{((d-1)!)^s} ((d-1) \cdot s + 1)^{s-2}$.

*Proof:* By Lemma 10, we have to count the number of bipartite trees over the two disjoint sets $L_H$ and $R_H$ of size $s$ and $(d-1) \cdots +1$. Since $H$ is a tree, then there are no cycles. Consequently, each one of the vertices in $L_H$ is connected to $d$ distinct vertices in $R_H$. Moreover, no two vertices in $L_H$ share more than 1 vertex in $R_H$. For each vertex $v_\ell \in L_H$, let $S_v$ be the set of the $d$ right-side vertices that $v_\ell$ is connected to and also let the cycle $C_{v_\ell}$ be a cycle that consists of the $d$ vertices of $S_v$.

Consider the graph $\hat{H} = \left\langle \hat{R}_H, \hat{E}_H \right\rangle$, which is given by connecting each cycle $C_{v_{\ell_1}}$ to $C_{v_{\ell_2}}$ using a common vertex $v_r$ if and only if $v_r$ is connected to both $v_{\ell_1}$ and $v_{\ell_2}$. The resulting graph $\hat{H}$ is a Husimi graph over $(d-1) \cdot s + 1$ vertices, where the number of such (labeled) graphs is $\frac{((d-1) \cdot s+1)!}{((d-1)!)^s \cdot s!} ((d-1) \cdot s + 1)^{s-2}$ [45].

Finally, each set $S_v$ is determined by the (labeled) vertex in $R_L$. Thus, we multiply by $s!$ the above expression. ∎

We are now able to prove the result.

Let $M$ be a maximum matching of $G$. Similarly to the proof of Theorem 1, the proof is based on counting the expected number of vertices in $R$ that are not part of $M$, and on the decomposition of $G$ into its connected components.

We count the expected number of connected components with $s$ left-side vertices and $q = (d-1) \cdot s + 1$ right-side vertices. By Lemma 9, the maximum matching size of each such connected component is exactly $s$. Thus, there are $q - s$ right-side vertices that are not in $M$.

Let $H$ be a bipartite graph $H = \langle L_H + R_H, E_H \rangle$, with degree $d$ for all vertices in $L_H$, where $|L_H| = s$ and $|R_H| = q$. The probability $P_s$ that $H$ is connected is given by $P_s = \frac{(d!)^s T_s^d}{q^{d \cdot s}}$.

The remainder of the proof is similar to the proof of Theorem 1. ∎

## APPENDIX B
## THE LAMBERT-$W$ FUNCTION

The Lambert-$W$ function, usually denoted by $W(\cdot)$, is given by the following implicit representation:

$$z = W(z) \cdot e^{W(z)},$$

where $z$ is a complex number [35].

For real valued arguments, i.e. $z$ is real valued, $W(z)$ has two real-valued branches: the principal branch, denoted by $W_0(\cdot)$ and the branch $W_{-1}(\cdot)$. Figure 8 shows the two real-valued branches. For instance, $W_0(-e^{-1}) = W_{-1}(-e^{-1}) = -1$ and $W_0(0) = 0$.

Note that the notation $W(\cdot)$ usually relates to the principle branch, i.e. $W_0(\cdot)$. Thus, although one would expect that for real-valued $z$, $W(z \cdot e^z) = z$, this is only the case for $z \geq -1$; in case $z < -1$, $W_{-1}(z \cdot e^z) = z \neq W(z \cdot e^z)$.